\documentclass{PoS}

\usepackage{amsmath}
\usepackage{amssymb}
\usepackage{graphicx,color,slashed,braket}
\usepackage{bbm} 
\usepackage{ifpdf}
\usepackage{comment}
\usepackage{multirow,bigdelim, caption}
\usepackage{cite}
\usepackage{comment}
\allowdisplaybreaks

\newcommand{\be}{\begin{equation}}
\newcommand{\ee}{\end{equation}}
\newcommand{\ba}{\begin{eqnarray}}
\newcommand{\ea}{\end{eqnarray}}
\newcommand{\nn}{\nonumber}
\newcommand{\lp}{\left(}
\newcommand{\rp}{\right)}
\newcommand{\ls}{\left[}
\newcommand{\rs}{\right]}
\def\rmi{{\rm i}}
\def\ib{{\bar \imath}}
\def\jb{{\bar \jmath}}
\newcommand{\Db}{\overline{D3}}
\newcommand{\A}{\mathcal{A}}

\title{Constructing the Supersymmetric anti-D3-brane action in KKLT}

\ShortTitle{Anti-D3-brane action in KKLT}

\author{\speaker{Niccol\`o Cribiori}
\\
         Institute for Theoretical Physics, TU Wien,\\
         Wiedner Hauptstrasse 8-10/136, A-1040 Vienna, Austria\\
        E-mail: \email{niccolo.cribiori@tuwien.ac.at}}

\abstract{The derivation of  the complete anti-D3-brane low energy effective action in KKLT is reviewed. All worldvolume fields are included, together with the background moduli. The result is recast into a manifest supersymmetric form in terms of the three independent functions of $\mathcal{N}=1$ supergravity in four dimensions: the K\"ahler potential, the superpotential and the gauge kinetic function. The latter differs from the expression one would expect by analogy with the D3-brane case.}

\FullConference{Corfu Summer Institute 2019 "School and Workshops on Elementary Particle Physics and Gravity" (CORFU2019)\\
		31 August - 25 September 2019\\
		Corf\`u, Greece}

\begin{document}

\section{Introduction and motivation}
According to experimental data \cite{Aghanim:2018eyx}, at present the universe is undergoing a period of accelerated expansions, which is compatible with the fact that dark energy constitutes roughly the 68\% of the total energy density. The assumption that dark energy is a cosmological constant is also in agreement with what observed. Its measured energy density is extremely small, $\mathcal{O}(10^{-122})$ in Planck units, but positive, and therefore the associated vacuum is de Sitter. In trying to understand de Sitter from a fundamental perspective, one can turn its attention to string theory. In this framework, a general recipe to construct de Sitter vacua has been proposed by Kachru, Kallosh, Linde and Trivedi (KKLT) \cite{Kachru:2003aw}.  However, to find concrete working models revealed itself to be challenging and, recently, the very existence of de Sitter solutions in quantum gravity has been questioned \cite{Danielsson:2018ztv, Obied:2018sgi}. Since at present there is no common agreement on a specific flaw in the KKLT proposal, it is important to keep investigating it in all of its details.

An anti-D3-brane is a central ingredient in the KKLT scenario. Its role is twofold: it provides a positive contribution to the vacuum energy, lifting the KKLT anti-de Sitter vacuum to de Sitter, and it breaks supersymmetry spontaneously.\footnote{It is important also to recall that a positive cosmological constant always breaks supersymmetry. This is compatible with the fact that no superpartners of the Standard Model particles have been observed in the colliders so far.} While it is known that D-branes in flat space preserve 16 supercharges linearly and 16 non-linearly, in a background with fluxes the situation is more complicated. Indeed, several works have been devoted to understand the breaking of supersymmetry by the anti-D3-brane in KKLT and the subsequent non-linear realization, see for example \cite{Kallosh:2014wsa, Bergshoeff:2015jxa, Kallosh:2015nia,Garcia-Etxebarria:2015lif,Bandos:2015xnf,Dasgupta:2016prs,Vercnocke:2016fbt,Kallosh:2016aep,Bandos:2016xyu,Aoki:2016tod,Aalsma:2017ulu,Kallosh:2017wnt,GarciadelMoral:2017vnz,Cribiori:2017laj,Krishnan:2018udc,Aalsma:2018pll,Cribiori:2018dlc}. The emerging picture suggests that the anti-D3-brane in KKLT can be seen as an instance of a more general scenario, called brane supersymmetry breaking \cite{Sugimoto:1999tx, Antoniadis:1999xk, Angelantonj:1999jh, Aldazabal:1999jr, Angelantonj:1999ms,Dudas:2000nv, Pradisi:2001yv}, in which a positive contribution to the scalar potential of the effective theory is induced by the presence of mutually non-BPS objects in the vacuum.

An essential tool in the investigation of the anti-D3-brane is the corresponding low energy effective theory. In this respect, it is important to recall that the action of a D-brane in a flux background is currently known up to quadratic order in the fermions \cite{Grana:2002tu, Grana:2003ek, Marolf:2003ye, Tripathy:2005hv, Martucci:2005rb, Bergshoeff:2013pia}. The anti-D3-brane action in KKLT has been analysed from several perspectives, with a particular care for the fermionic part \cite{Kallosh:2014wsa, Bergshoeff:2015jxa, Kallosh:2015nia,Garcia-Etxebarria:2015lif,Dasgupta:2016prs,GarciadelMoral:2017vnz}. Indeed, a simplified setup is obtained when placing the anti-D3-brane on top of the O3$^-$-plane, which is another ingredient of the KKLT background, since the orientifold projection removes all world volume bosons,  keeping only the fermions. A  step towards the derivation of the complete anti-D3-brane action has been performed recently, in \cite{GarciadelMoral:2017vnz}. In \cite{Cribiori:2019hod}, this result has been extended by considering the full supergravity background in KKLT and by including also the world volume scalar fields and the ${\rm U}(1)$ gauge vector. Here, the main steps of this work are summarized.

While the computation of the anti-D3-brane component form action is conceptually similar to the D3-brane case, its reformulation in a manifest four-dimensional supersymmetric language, namely superspace or tensor calculus, is more involved. This complication is related to the fact that on the anti-D3-brane world volume supersymmetry is non-linearly realized. As it is going to be reviewed, particular care is required for the gauge kinetic function. Indeed, a naive modification of the corresponding D3-brane expression would lead to an anti-holomorphic gauge kinetic function, which is not compatible with supersymmetry. However, by exploiting the features of non-linear supersymmetry, it is possible to write a holomorphic gauge kinetic function which reproduces correctly the anti-D3-brane component form action.

It would be interesting to repeat a similar study for other anti-Dp-branes. For example, based on \cite{Kallosh:2018nrk}, uplifting anti-D6-branes have been employed recently in a KKLT-inspired construction in type IIA \cite{Cribiori:2019bfx} and in the mass production procedure of \cite{Kallosh:2019zgd,Cribiori:2019drf,Cribiori:2019hrb}. Another interesting development would be to investigate the non-abelian case, namely to consider a stack of several (anti-)Dp-branes.

In this work, the conventions of \cite{Cribiori:2019hod} are followed. For the supergravity reformulation, these are the conventions of \cite{freedman2012supergravity}.

\section{The anti-D3-brane action in KKLT}
In this section, the main steps in the derivation of the anti-D3-brane action in KKLT are summarized. All world volume fields are included, namely six real scalars, four Weyl fermions and one vector,  together with their couplings to the background moduli fields, namely the axio-dilaton $\tau = C_0 + {\rm i} e^{-\phi}$, the single K\"ahler modulus $T$ and the complex structure moduli $U^A$. For convenience, the bosonic and fermionic parts of the action are discussed separately in sections \ref{sec:Sbos} and \ref{sec:Sferm}, respectively. The latter will contain up to quadratic terms in the world volume fermions, but terms with one world volume and one closed string fermion are not considered.

\subsection{Bosonic part}\label{sec:Sbos}

The bosonic part of the anti-D3-brane action in four-dimensional Einstein frame is
\begin{align}
S^{\Db}_{bos} &= S^{\rm DBI} + S^{\rm CS}\,,\\
S^{\rm DBI} &= - \int d^4x \sqrt{-\det \left({\rm P}\left[ g_{\mu\nu} + e^{-\frac{\phi}{2}} B_{\mu\nu}\right] + e^{-\frac{\phi}{2}} F_{\mu\nu} \right)}\,,\\
S^{\rm CS} &= -\int {\rm P} \left( (C_0 + C_2 + C_4) \wedge e^{B_2} \right) \wedge e^{F}\,,
\end{align}
where $B_2$ denotes the NSNS Kalb-Ramond field, $F_{\mu\nu }$ the field strength of the U$(1)$ gauge field living on the brane, $C_p$ are the RR potentials and P is the pullback to the brane world volume. The background of interest is warped \cite{Giddings:2001yu} and, in order to allow for a controlled uplift, a highly warped throat is needed where the anti-D3-brane is energetically attracted and sits. In this region, the Einstein frame metric is chosen to be
\begin{equation}
\label{gsw}
ds^2 = e^{2\A(z)-4u(x)}g_{\mu\nu} dx^\mu dx^\nu + e^{\frac43u(x)-\frac23 \A(z)}g_{a\bar b} dz^a d\bar z^{\bar b},
\end{equation}
where external indices are labeled by $\mu,\nu=0,1,2,3$,  internal complex indices by $a, \bar b=1,2,3$, $e^{-4\A(z)}$ is the warp factor and $e^{6u}=vol_6$ the volume of the compact manifold, whose dependence on Im$(T)$ is going to be specified below.\footnote{The metric \eqref{gsw} does not solve the mixed components of the ten-dimensional Einstein equations with one internal and one external index. To cure this problem, one has to introduce (derivative) compensator terms~\cite{Frey:2008xw}. These compensators will not affect the kinetic terms of the scalar fields in four-dimensions, which are fixed by the superconformal symmetry, before this is broken to obtain Poincar\'e supergravity. As explained in \cite{Martucci:2014ska,Martucci:2016pzt}, the superconformal symmetry relates the kinetic terms of the scalar fields to the normalization of the four-dimensional Ricci scalar, which does not contain derivatives and therefore it does not receive contributions from the compensators. However, it would be important to have an independent confirmation by doing a proper dimensional reduction of the anti-D3-brane action, extending the result of \cite{Cownden:2016hpf} for a supersymmetric D3-brane.}
In this background $B_2 =0= C_2$: the external components are projected out by the orientifold, while the internal and mixed ones vanish due to $h^{1,1}_-=0$ and $h^{1,0}=0=h^{0,1}$, respectively. 

Looking at the DBI-part of the action, the warp factor, the internal metric and the dilaton, via the pullback P, become functions of the world volume scalars $H^a$  indicating the position of the brane in the strongly warped region. For the rest of the discussion, small fluctuations around such a position are studied and are denoted with the same letter $H^a$ for convenience. Taking this into account, one has
\begin{equation}
\label{eq:DBIaction}
\begin{aligned}
S^{\rm DBI} = -  \int d^4x \sqrt{-g_4}\bigg( &e^{4 \mathcal{A}(H,\bar H)-8 u(x)} + \frac12 e^{\frac43\mathcal{A}(H,\bar H)-\frac83u(x)} g_{a\bar b}(H, \bar H) \partial_\mu H^a  \partial^\mu \bar H^{\bar b}\\
& +\frac{e^{-\phi(H, \bar H)}}{4} F_{\mu\nu} F^{\mu\nu} +\dots \bigg)\,,
\end{aligned}
\end{equation}
where dots denote subleading corrections.\footnote{Following \cite{McGuirk:2012sb}, the action is expanded in powers of the string length, which is eventually set to one.} The DBI-part of the action, which is the same for the D3-brane and the anti-D3-brane, encodes the information on the kinetic terms of the scalar fields. In four-dimensional $\mathcal{N}=1$ supergravity, the kinetic terms are contained in the K\"ahler potential, which in this setup depends on the open and closed string moduli \cite{DeWolfe:2002nn}
\begin{equation}
\label{eq:Kpot}
K =-2\log vol_6= -3 \log\left[-{\rm i} (T-\bar{T}) + k(H,\bar H) \right]=-12 u\,.
\end{equation}
Here,  $k(H,\bar H)$ is the K\"ahler potential corresponding to the internal Calabi-Yau metric. It does not break the no-scale structure and it is related to the metric $g_{a \bar b}$ by $\partial_{H^a}\partial_{\bar{H}^{\bar b}} k(H,\bar H) \approx\frac{1}{6} e^{\frac43 (\A+u)}g_{a\bar b}$, where subleading terms are neglected.

Recalling that in the background of interest $C_4 = \alpha(z,\bar z) \sqrt{-g_4}d^4x$ where $g_4$ is the determinant of the unwarped four-dimensional metric, the CS-part of the action reduces to
\begin{equation}
\begin{aligned}
\label{eq:CSaction}
S^{\rm CS} &=- \int \left(\frac12 C_0(H,\bar H)  F \wedge F + C_4 (H, \bar H)\right)\\
&= - \int d^4x \sqrt{-g_4}\left(-\frac{\text{Re}(\tau)}{8} \frac{\epsilon^{\mu\nu\rho\sigma}}{\sqrt{-g_4}}F_{\mu\nu}F_{\rho\sigma} + \alpha(H,\bar H)+\dots\right).
\end{aligned}
\end{equation}

The bosonic action of the anti-D3-brane is then obtained by putting together the two pieces \eqref{eq:DBIaction} and \eqref{eq:CSaction}. Notice in particular that the first term in the DBI-action~\eqref{eq:DBIaction} and the second term in the CS-action \eqref{eq:CSaction} combine in
\be
\label{Phipm}
\Phi_\pm \equiv e^{4 \A(H,\bar H)-8 u} \pm \alpha(H,\bar H),
\ee
the plus sign being for the anti-D3-brane and the minus for the D3-brane. The scalar potential for a D3-brane, $V_{D3}= \Phi_-$, vanishes because the DBI-part and the CS-part cancel against each other, on-shell. On the other hand, for an anti-D3-brane the contributions add up.  Hence, one has
\begin{equation}\label{eq:bosonic}
\begin{aligned}
S^{\Db}_{bos} &= S^{DBI} + S^{CS}\\
&=- \int d^4x \sqrt{-g_4}\bigg(\frac12 e^{\frac43\A(H,\bar H)-\frac83u(x)}g_{a\bar b}\partial_\mu H^a \partial^\mu \bar H^{\bar b} +V_{\Db}(H,\bar H)\\
&\qquad\qquad\qquad\qquad+ \frac{{\rm Im}(\tau)}{4}F_{\mu\nu}F^{\mu\nu} - \frac{{\rm Re}(\tau)}{8}\frac{\epsilon^{\mu\nu\rho\sigma}}{\sqrt{-g_4}}F_{\mu\nu}F_{\rho\sigma} +\dots\bigg),
\end{aligned}
\end{equation}
where the scalar potential is
\be
\label{eq:scalarpot}
V_{\Db}(H, \bar H) =  \Phi_+=2  e^{4 \A(H,\bar H)-8 u}.
\ee

In the background considered here, a D3-brane preserves linear $\mathcal{N}=1$ supersymmetry on its world volume. It is described by a holomorphic gauge kinetic function $f(\tau) =- {\rm i} \, \tau$, whose real part, Re$(f(\tau)) = {\rm Im}(\tau)=e^{-\phi}$, controls the coupling of the Maxwell term, while the imaginary part, Im$(f(\tau)) =- {\rm Re}(\tau)=-C_0$, the theta term. Instead, an anti-D3-brane has a sign difference in the CS-action, which would naively lead to an anti-holomorphic gauge kinetic function, $f(\bar \tau) = {\rm i}\, \bar \tau$, incompatible with the rules of supersymmetry. Therefore, the naive guess $f(\bar \tau)$ is not the correct gauge kinetic function describing an anti-D3-brane. The correct anti-D3-brane holomorphic gauge function will be constructed in section \ref{sec:gkf}, by exploiting the properties of non-linear supersymmetry.

\subsection{Fermionic part}
\label{sec:Sferm}
When supersymmetry is spontaneously broken, a Goldstino is present in the low energy spectrum. In the setup under consideration,  since the anti-D3-brane is the sole source of supersymmetry breaking, the Goldstino will be provided by one (combination) of its world volume fermions. In this perspective, the fermionic part of the anti-D3-brane action is therefore crucial in understanding the supersymmetry breaking mechanism in KKLT.

The four world volume fermions on the anti-D3-brane can be divided into $\lambda$, which is a singlet under the SU$(3)$ holonomy group of the internal manifold, and $\chi^i$, with $i=1,2,3,$ which transform as a triplet. This triplet of fermions is massive and its masses are controlled by the imaginary self dual (ISD) part
\be
G_3^{\rm ISD} = \frac12 (G_3 - {\rm i} *_6 G_3)
\ee
of the background flux $G_3=F_3-{\rm i} e^{-\phi}H_3$~\cite{McGuirk:2012sb, Bergshoeff:2015jxa}. Notice that the ISD condition allows for a non-vanishing primitive (2,1) and a (0,3) $G_3$ flux piece, which would give a mass to $\chi^i$ and to $\lambda$ respectively \cite{Bergshoeff:2015jxa}. Therefore, all fermions would be massive and there would be no Goldstino. However, as shown in \cite{Baumann:2010sx, Dymarsky:2010mf}, the (0,3) $G_3$ flux piece localizes on top of the D7-branes which are present in the bulk of the warped Calabi-Yau. As a consequence, the pull-back of this (0,3) $G_3$ flux onto the anti-D3-brane world volume vanishes, since the anti-D3-brane sits at the bottom of a warped throat. This means that the world volume fermion $\lambda$ is massless and can be identified as the Goldstino of the probe anti-D3-brane.

On the other hand, the presence of a non-vanishing (0,3) $G_3$ flux piece in the background breaks supersymmetry via a Gukov--Vafa--Witten superpotential~\cite{Gukov:1999ya}
\be
\label{WGVW}
W_{GVW} =\int G_3 \wedge  \Omega \neq 0\,,
\ee
where $\Omega$ is the (3,0) form of the Calabi-Yau. Indeed, this superpotential produces a non-vanishing F-term for the K\"ahler modulus $T$
\be
D_T W_{GVW} = K_T W_{GVW} \neq 0\,.
\ee
Moreover, besides a non-vanishing $W_{GVW}$, the KKLT background has also a non-perturbative superpotential term 
\be
\label{Wnp}
W_{np} = A e^{{\rm i} aT}\,,
\ee
which is needed to stabilize the K\"ahler modulus $T$ that otherwise would be a flat direction. Possible sources for such an additional contribution can be either gaugino condensation or Euclidean D3-brane instantons effects.\footnote{The effect of this non-perturbative contribution on the background as well as the anti-D3-brane uplift has recently received considerable attention~\cite{Moritz:2017xto, Moritz:2018sui, Kallosh:2018wme, Moritz:2018ani, Kallosh:2018psh, Gautason:2018gln, Hamada:2018qef, Kallosh:2019axr, Kallosh:2019oxv, Hamada:2019ack, Carta:2019rhx, Gautason:2019jwq,Kachru:2019dvo}. The central question is whether or not the gaugino condensation on a stack of D7-branes can be described in ten dimensions and  what the detailed backreaction of the gaugino condensate on the anti-D3-brane would be.} Notice that in principle $A$ is a function of the anti-D3-brane fields. However, since the gaugino condensation or Euclidean D3-brane instantons arise from the bulk Calabi-Yau bulk region, while the anti-D3-brane sits at the bottom of a highly warped throat, this dependence is expected to produce corrections which are highly suppressed compared to the tree-level potential in equation~\eqref{eq:scalarpot}.

After the non-perturbative effects are included, one can find a solution to the F-term equation
\be
D_T (W_{GVW} + W_{np})= 0\,, 
\ee
which gives
\begin{equation}
\label{KKLTW0}
W_{GWV} = - W_{np} - K_T^{-1}\partial_T W_{np}.
\end{equation}
This solution is a supersymmetric anti-de Sitter vacuum that is then uplifted by the anti-D3-brane to the KKLT de Sitter vacuum. 
The relation \eqref{KKLTW0} is central in KKLT. It establishes a delicate balance between perturbative and non-pertubative effects, leading to a supersymmetry restoration and to a stable anti-de Sitter vacuum.\footnote{See \cite{Linde:2020mdk} for a recent discussion on the possibility of having a KKLT uplift without an anti-de Sitter vacuum.} In particular, the perturbative superpotential $W_{GWV}$ is required to be small, in order to keep corrections under control. Evidence for the existence of vacua with $|W_{GWV}|\ll 1$ has been provided by means of statistical techniques (see \cite{Douglas:2006es} for a review) and an algorithm to construct explicit examples has been proposed recently in \cite{Demirtas:2019sip}.

Having clarified the breaking of supersymmetry by the anti-D3-brane in KKLT, one can complete the computation of the effective action. The part of the action that is quadratic in world volume fermions is given in Einstein frame by~\cite{Martucci:2005rb, Bergshoeff:2005yp, Bergshoeff:2015jxa}
\be
\begin{aligned}
\label{Sfer10d}
S^{\Db}_{fer} = 2\int &d^4x \sqrt{-g_4}\, \bigg[ e^{4\mathcal{A}-8 u}\bar\theta\Gamma^\mu \left(\nabla_\mu
-\frac{1}{4}e^\phi F_\mu \tilde \Gamma_{0123}\right)\theta\\
&+\frac{1}{8\cdot 4!} \bar \theta\left(e^{\frac{16\mathcal{A}}{3}-\frac{32u}{3}}\Gamma^{\mu mnpq}\, F_{\mu mnpq}-2e^{\frac{8\mathcal{A}}{3}-\frac{16u}{3}} \Gamma^{\mu\nu\rho mn} \,F_{\mu\nu\rho mn}\right)\tilde \Gamma_{0123}\theta\\
&-\frac{\rm i}{24}e^{6\mathcal{A}-12u + \frac{\phi}{2}}\,(G_{mnp}^{\rm ISD}-\bar G_{mnp}^{\rm ISD})\,\bar\theta\Gamma^{mnp}\theta\bigg]. 
\end{aligned}
\ee
Here, $\theta$ is a 16-components Majorana-Weyl spinor of type IIB theory, the indices $m,n,\ldots =4,5,\dots,9$ are internal, the matrices $\Gamma_{\ldots}$ have curved but unwarped indices, while $\tilde \Gamma_{\ldots}$ have flat indices. Following~\cite{Grana:2002tu,Bergshoeff:2015jxa}, one can decompose the spinor $\theta$ into four-dimensional Weyl spinors, namely an SU$(3)$-singlet  $P_L\lambda$ and a triplet $P_L \chi^i$. The dimensional reduction of the kinetic term and of the last line of \eqref{Sfer10d} was performed in \cite{Bergshoeff:2015jxa}, while for the remaining terms one can see \cite{Cribiori:2019hod}. Additional details are reported in the appendix \ref{appAferm}.  
Eventually, the action \eqref{Sfer10d} written in terms of four-dimensional fields reduces to
\begin{align}
S^{\Db}_{fer}  &=2 \int d^4 x \sqrt{-g_4}e^{4\mathcal{A}-8 u}\bigg[\bar\lambda P_R\gamma^\mu \nabla_\mu \lambda + \delta_{i\bar \jmath}\bar\chi^{\bar \jmath} P_R \gamma^\mu \nabla_\mu \chi^i \nonumber\\
&\qquad\qquad\qquad\qquad\qquad+\frac{\rm i}{4 {\rm Im}(\tau)} \partial_\mu {\rm Re}(\tau)\lp \bar\lambda P_R\gamma^\mu \lambda + \delta_{i\bar \jmath}\bar\chi^{\bar \jmath} P_R \gamma^\mu \chi^i \rp\nn\\
&\qquad\qquad\qquad\qquad\qquad-\frac{\rm i}{4 {\rm Im}(T)} \partial_\mu {\rm Re}(T)\lp 3\bar\lambda P_R\gamma^\mu \lambda - \delta_{i\bar \jmath}\bar\chi^{\bar \jmath} P_R \gamma^\mu \chi^i \rp\\
&\qquad\qquad\qquad\qquad\qquad+\frac{1}{12} \omega_\mu^{k\bar k} \delta_{k\bar k}\lp 3\bar\lambda P_R\gamma^\mu \lambda - \delta_{i\bar \jmath}\bar\chi^{\bar \jmath} P_R \gamma^\mu \chi^i \rp\nn\\
&\qquad\qquad\qquad\qquad\qquad+\left(
\frac12 m_{ij} \bar\chi^i P_L \chi^j+c.c.\right)\bigg]\nonumber\,,
\end{align}
where the fermionic mass depends on the $G^{\rm ISD}_3$ flux as
\begin{align}
\label{mij}
m_{ij} &= \frac{\sqrt 2}{8} \rmi e^{2\A-4u+\frac{\phi}{2}}(e^c_i e^d_j+e^c_j e^d_i)\Omega_{abc} g^{a\bar a}g^{b\bar b}\bar G^{\rm ISD}_{d\bar a\bar b}\,
\end{align}
and the K\"ahler modulus $T$ is defined as
\begin{equation}
T = \int_{\Sigma_4}\left(C_4-\frac{\rm i}2 J \wedge J\right).
\end{equation}

Eventually, the complete anti-D3-brane action in the KKLT background is
\be
\begin{aligned}
\label{SDbcomplete}
S^{\Db} &= S^{\Db}_{bos} + S^{\Db}_{fer}\\
&=- \int d^4x \sqrt{-g_4}\bigg(2  e^{4 \A-8 u}+\frac12 e^{\frac43 \A-\frac83 u}g_{a\bar b}\partial_\mu H^a \partial^\mu \bar H^{\bar b} \\
&\quad\qquad\qquad\qquad\quad+ \frac{{\rm Im}(\tau)}{4}F_{\mu\nu}F^{\mu\nu} - \frac{{\rm Re}(\tau)}{8}\frac{\epsilon^{\mu\nu\rho\sigma}}{\sqrt{-g_4}}F_{\mu\nu}F_{\rho\sigma}\bigg)\\
&+2 \int d^4x\sqrt{-g_4}e^{4\mathcal{A}-8 u}\bigg[\bar\lambda P_R\gamma^\mu \nabla_\mu \lambda + \delta_{i\bar \jmath}\bar\chi^{\bar \jmath} P_R \gamma^\mu \nabla_\mu \chi^i  \\
&\qquad\qquad\qquad\qquad\qquad+\frac{\rmi}{4 {\rm Im}(\tau)} \partial_\mu {\rm Re}(\tau)\lp \bar\lambda P_R\gamma^\mu \lambda + \delta_{i\bar \jmath}\bar\chi^{\bar \jmath} P_R \gamma^\mu \chi^i \rp\\
&\qquad\qquad\qquad\qquad\qquad-\frac{\rmi}{4 {\rm Im}(T)} \partial_\mu {\rm Re}(T)\lp 3\bar\lambda P_R\gamma^\mu \lambda - \delta_{i\bar \jmath}\bar\chi^{\bar \jmath} P_R \gamma^\mu \chi^i \rp\\
&\qquad\qquad\qquad\qquad\qquad+\frac{1}{12} \omega_\mu^{k\bar k} \delta_{k\bar k}\lp 3\bar\lambda P_R\gamma^\mu \lambda - \delta_{i\bar \jmath}\bar\chi^{\bar \jmath} P_R \gamma^\mu \chi^i \rp\\
&\qquad\qquad\qquad\qquad\qquad+\frac12 m_{ij} \bar\chi^i P_L \chi^j+\frac12 \overline m_{\bar \imath \bar \jmath} \bar\chi^{\bar \imath} P_R \chi^{\bar\jmath}\bigg]\,.
\end{aligned}
\ee

\subsection{Comparison with the D3-brane action}
\label{sec:D3comp}

It is instructive to compare the anti-D3-brane action \eqref{SDbcomplete} to the D3-brane action in the KKLT background. To this end, recall that the D3-brane differs from the anti-D3-brane by a sign flip of the RR fields. In practice, this amounts to a change in the sign of the terms involving $\partial_\mu{\rm Re}\tau$ and $\partial_\mu{\rm Re} T$, arising from $F_1$ and $F_5$ respectively, and to the replacement
\begin{equation}
\label{GIASD}
 G^\text{ISD}_3\to-G^\text{IASD}_3, \qquad  \text{where} \qquad G^\text{IASD}_3 = \frac12 (G_3 + {\rm i}*_6 G_3).
\end{equation}
In the background \cite{Giddings:2001yu}, and in KKLT as well, one has $G^{IASD}_3=0$ and therefore the fermionic mass term in \eqref{SDbcomplete}, after the substitution \eqref{GIASD}, vanishes for the D3-brane. In addition to this, also the uplifting term $2 e^{4\mathcal{A}-8u}$, which gives a positive contribution to the scalar potential in \eqref{SDbcomplete}, is not present in the D3-brane case, since the combination $\Phi_-$  in \eqref{Phipm} is identically zero, on-shell.  The remaining couplings are then described by a standard four-dimensional $\mathcal{N}=1$ supergravity action coupled to a single vector multiplet, containing $\lambda$ and $A_\mu$, and to three chiral multiplets, containing $H^a$ and $\chi^a \equiv e^a_i \chi^i$:
\begin{align}
\label{SDcomplete}
S^{D3} &= S^{D3}_{bos} + S^{D3}_{fer}\nn\\
&=- \int d^4x \sqrt{-g_4}\bigg(\frac12 e^{\frac43 \A-\frac83 u}g_{a\bar b}\partial_\mu H^a \partial^\mu \bar H^{\bar b} \nn\\
&\quad\qquad\qquad\qquad\quad+ \frac{{\rm Im}(\tau)}{4}F_{\mu\nu}F^{\mu\nu} + \frac{{\rm Re}(\tau)}{8}\frac{\epsilon^{\mu\nu\rho\sigma}}{\sqrt{-g_4}}F_{\mu\nu}F_{\rho\sigma}\bigg)\nn\\
&+2 \int d^4x\sqrt{-g_4}e^{4\mathcal{A}-8 u}\bigg[\bar\lambda P_R\gamma^\mu \nabla_\mu \lambda + \delta_{i\bar \jmath}\bar\chi^{\bar \jmath} P_R \gamma^\mu \nabla_\mu \chi^i  \\
&\qquad\qquad\qquad\qquad\qquad-\frac{\rmi}{4 {\rm Im}(\tau)} \partial_\mu {\rm Re}(\tau)\lp \bar\lambda P_R\gamma^\mu \lambda + \delta_{i\bar \jmath}\bar\chi^{\bar \jmath} P_R \gamma^\mu \chi^i \rp\nn\\
&\qquad\qquad\qquad\qquad\qquad+\frac{\rmi}{4 {\rm Im}(T)} \partial_\mu {\rm Re}(T)\lp 3\bar\lambda P_R\gamma^\mu \lambda - \delta_{i\bar \jmath}\bar\chi^{\bar \jmath} P_R \gamma^\mu \chi^i \rp\nn\\
&\qquad\qquad\qquad\qquad\qquad+\frac{1}{12} \omega_\mu^{k\bar k} \delta_{k\bar k}\lp 3\bar\lambda P_R\gamma^\mu \lambda - \delta_{i\bar \jmath}\bar\chi^{\bar \jmath} P_R \gamma^\mu \chi^i \rp\bigg]\nn\,.
\end{align}

Useful information on the form of the K\"ahler potential can be obtained by studying the derivative couplings involving (derivatives of) the closed string axions and the open string fermions, both for the D3-brane and the anti-D3-brane. A detailed discussion can be found in \cite{Cribiori:2019hod}, but an interesting result is reported also here. In the component action \eqref{SDcomplete}, the couplings of $\lambda$ and $\chi^i$ to $\partial_\mu {\rm Re}(T)$ and to $\partial_\mu {\rm Re}(U^A)$, which is encoded in the spin connection term in the last line, have the same relative sign between the bilnear in $\lambda$ and in $\chi^i$. Actually, once the spin connection is expressed in terms of $\partial_\mu {\rm Re}(T)$, as in \cite{Cribiori:2019hod}, those couplings have exactly the same numerical coefficient. This implies that the complex structure sector and the K\"ahler sector have to couple to the chiral multiplets on the D3-brane in the same way. The K\"ahler potential which is correctly reproducing these interactions is therefore
\ba\label{eq:D3Kahler}
K &=& -\log\ls-\rmi (\tau-\bar{\tau})\rs -3\log\ls-\rmi (T-\bar{T})\lp-\rmi \int \Omega \wedge \bar \Omega \rp^{\frac13} + k(H,\bar H)\rs\cr
&=& -\log\ls-\rmi (\tau-\bar{\tau})\rs -\log\ls-\rmi \int \Omega \wedge \bar \Omega\rs \cr
&&\qquad\qquad\qquad\qquad-3\log\ls-\rmi (T-\bar{T}) + \frac{k(H,\bar H)}{\lp-\rmi \int \Omega \wedge \bar \Omega \rp^{\frac13}}\rs
\ea
and it couples the world volume scalars $H^a$ to the complex structure moduli $U^A$.
This is crucially different from the K\"ahler potential
\be
K = -\log\ls-\rmi (\tau-\bar{\tau})\rs -\log\ls-\rmi \int \Omega \wedge \bar \Omega\rs -3\log\ls- \rmi(T-\bar{T})\ + k(H,\bar H)\rs\,,
\ee
which one might expect but which does not seem to be compatible with \eqref{SDcomplete}. Finally, in order to still reproduce the kinetic term for the scalars $H^a$, the K\"ahler potential $k(H,\bar H)$ needs to be chosen such that $\partial_{H^a}\partial_{\bar{H}^{\bar b}} k(H,\bar H) \approx \frac{1}{6} e^{\frac43 (\A+u)}\lp-\rmi \int \Omega \wedge \bar \Omega \rp^{\frac13}g_{a\bar b}$.

\section{Supersymmetric formulation}

In this section, it is reviewed how the anti-D3-brane \eqref{SDbcomplete} action can be organized in a manifestly supersymmetric language in four-dimensional $\mathcal{N}=1$ supergravity. All couplings of minimal supergravity in four dimensions are encoded into three independent functions: the K\"ahler potential $K$, the superpotential $W$ and the gauge kinetic function $f$. The first is a real function of chiral multiplets, while the others are holomorphic. 

To describe the world volume fields of a D3-brane, one needs three chiral multiplets, containing the ${\rm SU}(3)$-triplet of fermions and the scalars, and a vector multiplet, encoding the ${\rm U}(1)$ gauge vector and the gaugino. For the anti-D3-brane, due to the underlying non-linear realization of supersymmetry, one has in general more freedom in the choice of the supersymmetric embedding. In the following, a consistent supersymmetric formulation of the anti-D3-brane action is constructed, drawing intuition from the D3-brane case. However, it is important to keep in mind that this construction is by no means unique and that the analogy with linear supersymmetry does not extend too much further. Indeed, the gauge kinetic function is a notable example in which this analogy between the two systems does not apply directly.

\subsection{From linear to non-linear supersymmetry}

Among the main ingredients of $\mathcal{N}=1$ linear supersymmetry in four-dimensions there are chiral and vector multiplets. The chiral multiplet is an irreducible representation of supersymmetry containing one complex scalar $\Phi$, one Weyl fermion  $P_L \Omega^\Phi$ and one complex auxiliary field $F^\Phi$,
\begin{equation}
\Phi = \{\Phi, P_L \Omega^\Phi, F^\Phi\}.
\end{equation}
The vector multiplet is a real scalar multiplet containing a vector field and its fermionic superpartner, the gaugino $P_L \Lambda$. Since the vector multiplet is not gauge invariant, it is usually more convenient to work with the associated gauge-invariant field strength multiplet. This is a chiral multiplet with a fermionic spacetime index 
\begin{equation}
P_L\Lambda_\alpha = \{P_L\Lambda_\alpha, (P_L \chi)_{\beta \alpha}, F_\alpha^\Lambda\},
\end{equation}
where
\begin{align}
(P_L\chi)_{\beta\alpha} &= \sqrt 2\left[-\frac14 (P_L\gamma^{ab}C)_{\beta\alpha}\hat F_{ab} + \frac i2 {\rm D} (P_LC)_{\beta\alpha}\right],\\
F_\alpha^\Lambda &= (\slashed{\mathcal{D}}P_R\Lambda)_\alpha.
\end{align}
It contains (the field strength of) the vector, the gaugino and  a real auxiliary field D. The matrix $C_{\alpha \beta}$, satisfying $C^T = - C$, is used to raise and lower fermionic indices, while $\hat F_{ab} =  e_a^\mu e_b^\nu (2\partial_{[\mu} A_{\nu]} + \bar \psi_{[\mu }\gamma_{\nu]}\lambda)$ is the covariant vector field strength.\footnote{The multiplet $P_L\Lambda_\alpha$ is the analogous of the chiral superfield $W_\alpha = -\frac14 \bar D^2 D_\alpha V$ defined in superspace.}

These multiplets are linear representations of supersymmetry, meaning that there exist supersymmetry transformations among their components which act linearly on the fields. Such a type of transformations does not exist for the world volume fields of the anti-D3-brane. Nevertheless, one can still write transformations for the anti-D3-brane world volume fields closing the supersymmetry algebra, but these will have in general a more complicated form, with a non-linear action on fields. An important point is that these more complicated transformations can be conveniently and efficiently constructed from the known linear ones, as it is going to be reviewed in the following. The central idea is to impose, on top of known multiplets, supersymmetric constraints admitting a non-linear solution. However, one might wonder how general is such a procedure of constructing non-linear supersymmetry representations from linear ones by means of constrained multiplets. In this respect, an important result is the one proved in \cite{DallAgata:2016yof, Cribiori:2017ngp}, in which it is shown that by using a specific set of constrained multiplets, one can describe any possible model with non-linear supersymmetry.

Constrained multiplets (or superfields) are a convenient approach to non-linear supersymmetry (see \cite{Cribiori:2019cgz} for a review). In such an approach, the non-linearity is implemented by imposing constraints on standard, linear, supersymmetric multiplets. These additional constraints have to be supersymmetric, in order to avoid an explicit breaking of supersymmetry. For example, an important constraint is the nilpotent constraint imposed on a chiral multiplet $X$ \cite{Rocek:1978nb,Lindstrom:1979kq,Casalbuoni:1988xh}
\begin{equation}
\label{X2=0}
X^2 = 0 \qquad \Longleftrightarrow \qquad X = \left\{\frac{\bar \Omega P_L \Omega}{2F}, P_L \Omega, F\right\}.
\end{equation}
The constraint is supersymmetric, therefore its solution is still a (chiral) multiplet. However, one can see that the scalar in the lowest component is replaced by a bilinear in the fermion. This is precisely the way in which non-linear supersymmetry is implemented. Indeed, while in the case of the standard, unconstrained, chiral multiplet supersymmetry transformations are mapping the fermion to the scalar, in the nilpotent chiral multiplet the supersymmetry transformations are mapping the fermion to the composite expression $\bar \Omega P_L \Omega/2F$. These supersymmetry transformations are thus non-llinear with respect to the component fields. Moreover, it is crucial to notice that the auxiliary field $F$ appears in the denominator. This means that, for the consistency of the non-linear solution, $F$ has to be non-vanishing and therefore supersymmetry is broken spontaneously in the vacuum by an F-term. As a consequence, the fermion $P_L\Omega$ transforms under supersymmetry with a shift proportional to the vacuum-expectation-value of $F$. This fermion is then a Goldstino, or it contributes to the Goldstino in the case in which other non-vanshing F-terms are present. In \cite{Cribiori:2017ngp}, it is shown that one can employ the nilpotent chiral multiplet \eqref{X2=0} to describe the Goldstino of any low energy effective theory with spontaneously broken supersymmetry. Therefore, this constrained multiplet will be used to describe the Goldstino of the anti-D3-brane.

The lesson is completely general: when imposing supersymmetric constraints on multiplets, one is in fact removing some of their components and replacing them with composite expressions of the remaining fields. As a consequence, supersymmetry transformations will act non-linearly on these remaining fields, but they will still close the supersymmetry algebra. In general, depending on the form of the constraint, a different component field can be removed. An organizing principle explaining the origin of the constraints has been proposed in \cite{DallAgata:2016yof}, where it has been shown that, given any generic multiplet $Q$, its lowest component field can be removed by imposing
\begin{equation}
\label{Qconstr}
X \bar X Q =0,
\end{equation}
where $X$ is the nilpotent Goldstino chiral multiplet. Therefore, the constraints \eqref{X2=0} and \eqref{Qconstr} are the only two ingredients one needs to construct non-linear supersymmetry representations out of linear ones and to describe any low energy effective theory with spontaneously broken supersymmetry, as for example the anti-D3-brane action in KKLT given in \eqref{SDbcomplete}. In particular, without loss of generality the Goldstino sector can be described by a nilpotent chiral multiplet, while all the remaining fields can be embedded into different linear multiplets $Q$ and, on each of them, the constraint \eqref{Qconstr} can be imposed. 

However, one has to keep in mind that this procedure of constructing non-linear supersymmetry from linear supersymmetry is just an efficient tool, but the analogy between linear and non-linear multiplets is more an artefact of the construction, rather than a fundamental aspect. Indeed, when supersymmetry is non-linearly realized, the structure of the multiplets is generically lost and, for example, there is no meaning in thinking of the gaugino as the would be linear superpartner of the vector.

The world volume fields of a D3-brane are six real scalars, four Weyl fermions and one gauge vector. These are organized into three chiral multiplets (three complex scalars and three Weyl fermions) and one vector multiplet (one fermion and one vector), or equivalently its field strength multiplet. The world volume field of the anti-D3-brane are the same, but there is no unique way to embed them into non-linear representations of supersymmetry. One possibility is to embed each of them into a different multiplet, where all the other components but the field itself are removed by means of constraints.  For the reasons explained above, a natural choice is to embed the Goldstino into a nilpotent chiral multiplet $X$. Then, the three remaining fermions can be described by three other chiral multiplets $Y^i$, subjected to the constraints $XY^i=0$, which remove the scalars in the lowest component. As in the linear case, the vector field can be embedded into a chiral field strength multiplet $P_L \Lambda$,  with the addition of the constraint $X P_L \Lambda=0$, removing the gaugino. Finally, the three complex scalars can be viewed as lowest components of three further chiral multiplets $H^a$, where the constraints $X\bar H^a =\text{chiral}$ remove their fermionic superpartners. Notice that all of these constraints descend from the general constraint \eqref{Qconstr} (the factor $\bar X$ is not needed when dealing with chiral multiplets only).\footnote{The constraint  $X\bar H^a =\text{chiral}$ is actually different from \eqref{Qconstr}. However, in \cite{DallAgata:2016yof} it is shown how such a constraint is equivalent to imposing two constraints of the general form $X\bar X Q=0$.} The reader is referred to \cite{DallAgata:2016yof,Cribiori:2019hod} for more details on these constraints and for their explicit solutions. The important information to be remembered here is that, modulo auxiliary fields, the constrained $X$ contains only the Goldstino, each of the constrained $Y^i$ contains one fermion, the constrained $P_L \Lambda$ contains only the vector and the constrained $H^a$ contain only the scalars in their lowest component. This is summarized in table \ref{tableconstr}.

\begin{table}[h!]
\begin{center}
\begin{tabular}{c|c|c|c}
world volume field & D3-brane & linear SUSY & non-linear SUSY ($\overline{D3}$)\\
\hline
 1 fermion (singlet) &\rdelim\}{1.7}{2cm}[vector mult.]  & $X$ & $X^2=0$ \\ 
 1 vector &  & $P_L\Lambda$ & $XP_L\Lambda=0$\\  
 3 fermions (triplet) & \rdelim\}{1.7}{2cm}[chiral mult.] & $Y^i$ &   $XY^i=0$  \\
 3 scalars & & $H^a$ & $X \bar H^a = {\text chiral}$ 
\end{tabular}
\caption{The embedding of the world volume fields into chiral multiplets. In the second column, it is reported how fields are paired in the supersymmetric case of a D3-brane. In the third column, a possible choice of chiral multiplets is given to describe the anti-D3-brane. These contains more components than needed, since they are linear representations of supersymmetry. The non-linear realization is implemented on them with the constraints in the fourth column, which leave only the desired world volume field (first column) as independent component.}
\label{tableconstr}
\end{center}
\end{table}

Having understood how to describe each of the world volume fields of the anti-D3-brane with supersymmetric but constrained multiplet, one can proceed with the construction of the supergravity action.

\subsection{The supergravity action}

In the superconformal approach \cite{freedman2012supergravity}, the general form of the $\mathcal{N}=1$ supergravity action in four dimensions is given by 
\begin{equation}
\label{SugraS}
S = [-3 X^0 \bar X^0 e^{-\frac{K}{3}}]_D + [(X^0)^3W]_F +  [f_{\Db}\, \bar \Lambda P_L \Lambda]_F,
\end{equation}
where $X^0$ is a compensator chiral multiplet associated to the old-minimal off-shell formulation. This multiplet has to be fixed to the four-dimensional Planck mass, $X^0 \equiv \kappa^{-1}e^\frac{K}{6}$,  in order to break the superconformal symmetry and obtain Poincar\'e supergravity. The action is completely specified by three independent functions: the K\"ahler potential $K$, the superpotential $W$ and the gauge kinetic function $f_{\Db}$. The first is a real function, while the others are holomorphic. In the following, the strategy to deduce the correct $K$, $W$ and $f_{\Db}$ for the anti-D3-brane in KKLT is outlined.

In order to reproduce the anti-D3-brane action \eqref{SDbcomplete}, a convenient starting point is to consider the following K\"ahler potential and superpotential
\begin{align}
K &= -3 \log \left(A - B X \bar X - C \delta_{i\bar\jmath} Y^i \bar Y^{\bar \jmath} \right),\\
W &= W_{GWV} + M^2 X + W_{np} + \frac12 \hat M_{ij} Y^i Y^j,
\end{align}
where the chiral multiplets $X$ and $Y^i$ are constrained as in table \ref{tableconstr}. The quantities $A$, $B$, $C$ and $\hat M_{ij}$ (and also $W_{GWV}$, $W_{np}$) are treated as constants for the time being. \footnote{Here, $A$ has not to be confused with the parameter in $W_{np}$ in \eqref{Wnp}, which will never appear explicitly in the following formulae.} They will acquire a field dependence when matching the supergravity action with \eqref{SDbcomplete}. The parameter $M^2$ has to be non-vanishing, since it is related to the supersymmetry breaking scale and eventually to the tension of the anti-D3-brane. In the conventions used in the previous section, it is given  by $M^2 = \sqrt 2$. Taking into account these expressions for $K$ and $W$, the relevant terms in the component expansion of \eqref{SugraS} are
\begin{equation}
e^{-1}\mathcal{L} = -\left(\frac{3B}{A}\right)\bar \Omega P_L {\slashed{\mathcal{D}}}\Omega - \left(\frac{3C}{A}\right)\delta_{i\bar \jmath} \bar\Omega^i P_L{\slashed{\mathcal{D}}}\Omega^{\bar \jmath} - \frac12 e^\frac{K}{2}\left(\hat M_{ij}\bar \Omega^{i}P_L  \Omega^j+c.c.\right)- \mathcal{V}+\dots,
\end{equation}
where the covariant derivatives on the fermions contain the spin connection, the composite K\"ahler connection and the Christoffel connection. The supergravity scalar potential, omitting D-terms, is given by
\begin{equation}
\mathcal{V} = V_{\Db} - 3\frac{|W|^2}{A^3},\qquad \text{where} \qquad V_{\Db} = \frac13 \frac{M^4}{A^2 B}
\end{equation}
is the anti-D3-brane contribution. Notice that to normalize the fermions $P_L\Omega^i$ as in \eqref{SDbcomplete}, one has to perform a field redefinition of the type
\begin{equation}
\label{fermredef}
P_L \Omega^i = {\rm i} M^2 \sqrt{\frac{A}{3C}} e^{2\mathcal{A}-4u}P_L \chi^i+\dots,
\end{equation}
where dots stand for higher order fermionic terms. A similar redefinition, with $C\to B$ and $P_L\chi^i \to P_L \lambda$, might be considered also for the singlet fermion $P_L\Omega$, but in practice this will not be necessary.

\subsubsection{K\"ahler potential}

In order to obtain the complete form of the K\"ahler potential for the anti-D3-brane, the first step is to embed the Goldstino $P_L\lambda$ and the triplet of fermions $P_L\chi^i$ into, respectively, a chiral multiplet $X$ and a triplet of chiral multiplets $Y^i$, satisfying the constraints in table \ref{tableconstr}. Then, one can make an educated guess for the field dependent coefficients $A$, $B$ and $C$
\begin{align}
\label{coeffA}
A &= (-{\rm i}(\tau-\bar \tau))^\frac13  f(U^A, \bar U^A) ( -{\rm i}(T-\bar T)),\\
\label{coeffB}
B &= \frac13 e^{-4\mathcal{A}}(-{\rm i}(\tau-\bar \tau))^{-\frac23} f(U^A, \bar U^A)^{\frac{1+\alpha_1}{3}}  (-{\rm i}(T-\bar T))^{1+\alpha_2} ,\\
\label{coeffC}
C &= \frac13 e^{-4\mathcal{A}}(-{\rm i}(\tau-\bar \tau))^{-\frac23}  f(U^A, \bar U^A)^{\frac{1+\beta_1}{3}}  (-{\rm i}(T-\bar T))^{1+\beta_2}.
\end{align}
From now on, the notation $f(U^A, \bar U^A)=-\rmi \int \Omega \wedge \bar \Omega\,$ is used to avoid confusion between the holomorphic (3,0)-form and the fermions $P_L\Omega$, $P_L\Omega^i$. In the expressions \eqref{coeffA}, \eqref{coeffB} and \eqref{coeffC}, the powers of $\tau$ are fixed by modular invariance. The couplings to the K\"ahler and complex structure moduli, however, are still to be determined. The couplings of these moduli to $X$ are fixed by comparing the supergravity scalar potential with the scalar potential in~\eqref{SDbcomplete}, while those to $Y^i$ are fixed by matching with the K\"ahler covariant kinetic terms of the massive world volume fermions. By direct computation, one finds
\begin{equation}
\alpha_1 = -3,\qquad \alpha_2 = -1, \qquad \beta_1 = -1, \qquad \beta_2 =-2.  
\end{equation}
Therefore, the field dependent coefficients $A$, $B$ and $C$ have been completely fixed and the K\"ahler potential for the all fields, but the world volume scalars $H^a$, is given by
\begin{equation}
\begin{aligned}\label{GMPQZ}
K=&-\log(-{\rm i}(\tau-\bar{\tau}))-3\log\left[ (-\rmi (T-\bar T))f(U^A,\bar{U}^A)^\frac13\right]\\
&-3\log\left(1-\frac{e^{-4{\cal A}}X\bar{X}}{3(-{\rm i}(\tau-\bar{\tau}))(-\rmi(T-\bar T))f(U^A,\bar U^A)}\right.\\
&\left.\qquad\qquad-\frac{e^{-4{\cal A}}\delta_{i\bar\jmath}Y^i\bar{Y}^{\bar \jmath}}{3(-{\rm i}(\tau-\bar{\tau}))(-{\rm i} (T-\bar T))^2f(U^A,\bar{U}^A)^\frac13}\right).
\end{aligned}
\end{equation}
The field redefinition \eqref{fermredef} between the fermions $P_L\Omega^i$ inside $Y^i$ and $\chi^i$ becomes then
\begin{align}
\label{fermionredef}
P_L\Omega^i &= {\rm i} M^2e^{4\mathcal{A}}(-{\rm i}(\tau-\bar\tau))^\frac12  f(U^A,\bar{U}^A)^{\frac16}P_L \chi^i + \dots.
\end{align}
Notice that the terms involving $P_L\lambda$ have not been matched with $P_L\Omega$. This is due to the fact that $P_L\lambda$ is the Goldstino of the anti-D3-brane and, as such, is not physical. Indeed, in supergravity, the Goldstino is a pure gauge degree of freedom and can be set to zero by an appropriate unitary gauge choice. For this reason, it is not really needed to match precisely $P_L \lambda$ with $P_L \Omega$, but it is sufficient to keep in mind that the Goldstino resides in the constrained chiral multiplet $X$ and thus $P_L \Omega \sim P_L \lambda +\dots$, where dots stand for higher order terms. The situation is crucially different in the D3-brane case, where all fermions are physical.

Inspired by the observation at the end of section \ref{sec:D3comp}, a prescription is now given to introduce in the K\"ahler potential \eqref{GMPQZ} the dependence on the world volume scalars $H^a$. First, one has to embed the scalar fields into the constrained multiplets $H^a$, such that $X\bar H^a=\text{chiral}$. These multiplets contain only a scalar in the lowest component as independent degree of freedom. Then, one lets $\mathcal{A}$ depend on $H^a$ and formally shifts the volume modulus as  $-{\rm i}(T-\bar{T})\to -{\rm i}(T-\bar{T})+k(H^a,\bar{H}^a)/f(U^A,\bar U^A)^{\frac13}=e^{4u}$. As a result, the K\"ahler potential of $\mathcal{N}=1$ supergravity in four dimensions becomes
\be
\begin{aligned}
K=&-\log(-{\rm i}(\tau-\bar{\tau}))-3\log\left[ (-\rmi (T-\bar T))f(U^A,\bar{U}^A)^\frac13+k(H^a,\bar H^a)\right]\\
&-3\log\left(1-\frac{e^{-4{\cal A}(H^a,\bar H^a)-4u}}{3(-{\rm i}(\tau-\bar{\tau}))f(U^A,\bar U^A)}X\bar{X}
-\frac{e^{-4{\cal A}(H^a,\bar H^a)-8u}}{3(-{\rm i}(\tau-\bar{\tau}))f(U^A,\bar U^A)^\frac13}\delta_{i\bar\jmath}Y^i\bar{Y}^{\bar \jmath}\right).
\end{aligned}
\ee
In principle, also the superpotential $W_{np}$ would get a dependence on the scalars $H^a$, but this effect would be subleading in the present regime. Therefore, one can still consider \eqref{WSUGRA} as the correct expression for the superpotential in supergravity.

\subsubsection{Gauge kinetic function}
\label{sec:gkf}

Similarly to what done for the other fields, one has first to embed the world volume vector into a chiral field strength multiplet $P_L \Lambda_\alpha$ , which is then constrained as reported in table \ref{tableconstr}. The role of this constraint, namely $XP_L \Lambda=0$, is to remove the gaugino and leave only the U$(1)$ vector as independent physical degrees of freedom. Then, one can consider the standard supergravity action for the vector multiplet, which is \cite{freedman2012supergravity}
\be
\begin{aligned}
\label{LV}
-\frac14 [f(\tau) \bar \Lambda P_L \Lambda]_F &= \int d^4x\, \sqrt{-g_4}\left(-\frac{{\rm Re}(f)}{4} F_{\mu\nu}F^{\mu\nu} + \frac{{\rm Im}(f)}{8}\frac{\epsilon^{\mu\nu\rho\sigma}}{\sqrt{-g_4}} F_{\mu\nu}F_{\rho\sigma}+\dots\right),
\end{aligned}
\ee
with $f(\tau) = -{\rm i}\tau$ and where the dots stand for fermionic and auxiliary terms. Notice that the term proportional to ${\rm Im}(f) = - {\rm Re}(\tau)$ has the opposite sign compared to~\eqref{SDbcomplete}. This sign difference is due to the RR sign flip in the anti-D3-brane action. Therefore, in order to describe the anti-D3-brane in supergravity, one might naively try to use a gauge kinetic function depending on $\bar \tau$, namely an anti-holomorphic gauge kinetic function. This is not compatible with the rules of supersymmetry, since the gauge kinetic function has to be holomorphic. In \cite{Cribiori:2019hod}, it is shown explicitly how to deform the action \eqref{LV} in a supersymmetric way in order to obtain the desired sign flip. Here, the same result is presented from a slightly different perspective.

From the previous observation, one can conclude that the naive guess $ \bar f(\bar \tau)$ is not the correct anti-D3-brane gauge kinetic function. However, there is no reason to believe that it is not possible to embed consistently the coupling of the axio-dilaton to the vector on the anti-D3-brane in a supersymmetric way.  Indeed, the correct gauge kinetic function can be constructed by taking advantage of non-linear supersymmetry. In this respect, the idea is to find an appropriate constraint which transforms anti-chiral multiplets into chiral ones. In fact, one can even find a more powerful constraint, turning a generic multiplet, which does not need to be anti-chiral, into chiral. This constraint has been found in \cite{Cribiori:2019hod}. Given a generic superfield $Q$, in global supersymmetry one can construct
\begin{equation}
\hat Q =  \bar D^2 \left(\frac{\bar X }{\bar D^2 \bar X}Q\right),
\end{equation}
where $X$ is the nilpotent Goldstino chiral superfield and $\bar D^2$ is the chiral projector in superspace. The composite superfield $\hat Q$ is manifestly chiral and its lowest component is
\begin{equation}
\label{hatQQ}
\hat Q| = Q + \text{fermions}.
\end{equation}
The generalization of this construction to the superconformal setup is straightforward. One needs to introduce appropriate powers of the compensator multiplet $X^0$ and, by denoting with $\Sigma$ the chiral projector (see \cite{Cribiori:2019hod} and the references therein for more details on this operator), the desired expression is 
\begin{equation}
\hat Q = \Sigma\left(\frac{\bar X^0 e^{-\frac K6}\bar X }{\Sigma (\bar X^0 e^{-\frac K6}\bar X)}Q\right).
\end{equation}
Then, it is easy to see that the composite multiplet $\hat Q$ solves the constraint
\begin{equation}
X \bar X \hat Q = X \bar X Q.
\end{equation}
Notice that, for the definition of $\hat Q$ to be consistent, the F-term of $X$ has to be non-vanishing. This is indeed the case for the anti-D3-brane, since the multiplet $X$ provides the Goldstino. In addition, all the fermionic terms on the left hand side of \eqref{hatQQ} contain at least one Goldstino. Therefore, in the unitary gauge, this expression simplifies drastically to $\hat Q| = Q$.

One can now apply directly this construction to the problem at hand. Indeed, starting from the antichiral gauge kinetic function $\bar f (\bar \tau)$, one can construct the chiral expression
\be
\label{fD3b}
\hat f_{\Db} = \Sigma \left( \frac{\left(\bar X^0 e^{-\frac K6}\bar X\right)}{\Sigma \left(\bar X^0 e^{-\frac K6}\bar X \right)} \bar f(\bar \tau)\right).
\ee
This is a chiral multiplet that is anti-holomorphic in $\tau$ and that contains Goldstino interactions, implementing the non-linear realization of supersymmetry. In the unitary gauge, in which the Goldstino is set to zero, its lowest component is given by $\hat f_{\Db}| = \bar f(\bar \tau)$. Therefore, this quantity describes the same physical degrees of freedom as $\bar f(\bar \tau)$ and, at the same time, it can be consistently used into a supergravity action.

The couplings of the vector field of the anti-D3-brane are then described in supergravity by (in the unitary gauge)
\begin{equation}
\begin{aligned}
-\frac14 [\hat f_{\Db}(\bar \tau, \bar X) \bar \Lambda P_L \Lambda]_F &= \int d^4x\, \sqrt{-g_4}\left(-\frac{{\rm Re}(f)}{4} F_{\mu\nu}F^{\mu\nu} - \frac{{\rm Im}(f)}{8}\frac{\epsilon^{\mu\nu\rho\sigma}}{\sqrt{-g_4}} F_{\mu\nu}F_{\rho\sigma}+\dots\right)
\end{aligned}
\end{equation}
which reproduces precisely the terms in \eqref{SDbcomplete}.

\subsubsection{Superpotential}

As already anticipated, the superpotential describing the anti-D3-brane is
\be\label{WSUGRA}
W =  W_{GVW} + W_{np} + M^2 X + \frac12 \hat M_{ij}Y^i Y^j,
\ee
where the field dependent coefficient $\hat M_{ij}$, which is needed in order to give a mass to the triplet of fermions $P_L \Omega^i \sim P_L \chi^i + \dots$, has still to be determined. By looking at the explicit expression of the fermionic mass term in \eqref{mij}, one can see that it depends on $\bar \tau$. Therefore, this mass term cannot be inserted directly into the superpotential, because the latter has to be holomorphic. However, one can apply the lesson learned from the discussion on the gauge kinetic function and construct a chiral multiplet $\hat M_{ij}$ from the anti-chiral $m_{ij}$. The proper expression is
\be\label{eq:MIJ}
\hat M_{ij} =\Sigma \left(\frac{\left(\bar X^0 e^{-\frac K6}\bar X\right)e^{-4\A-2u} \,m_{ij}}{\,\Sigma\left(\bar X^0 e^{-\frac K6}\bar X\right)(-\rmi(\tau-\bar\tau))^\frac12 f(U^A,\bar U^A)^{-\frac16}}\right)
\ee
and it shares properties similar to the analogous construction $\hat f_{\Db}$. In particular, its lowest component in the unitary gauge is
\begin{equation}
\hat M_{ij}| = e^{-4\mathcal{A}-2u}(-{\rm i}(\tau-\bar\tau))^{-\frac12}f(U^A,\bar U^A)^{\frac16} m_{ij}.
\end{equation}
Having determined the explicit expression of $\hat M_{ij}$, the superpotential associated the anti-D3-brane action \eqref{SDbcomplete} is completely determined.

\section{Summary of the result and comment on ${\rm SL}(2, \mathbb{R})$ invariance}

The supergravity action of the anti-D3-brane is given by the following K\"ahler potential, superpotential and gauge kinetic function
\begin{align}
K=&-\log(-{\rm i}(\tau-\bar{\tau}))-3\log\left[ (-\rmi (T-\bar T))f(U^A,\bar{U}^A)^\frac13+k(H^a,\bar H^a)\right]\\
&-3\log\left(1-c_1 X \bar X - c_2 \delta_{i\jb} Y^i \bar Y^\jb\right),\nonumber\\
W &= W_{GVW} + W_{np} + M^2 X + \frac12 \hat M_{ij}Y^i Y^j,\\
\hat f_{\Db} &= \Sigma \left( \frac{\left(\bar X^0 e^{-\frac K6}\bar X\right)}{\Sigma \left(\bar X^0 e^{-\frac K6}\bar X \right)} \bar f(\bar \tau)\right),
\end{align}
where
\begin{align}
c_1 &= \frac{e^{-4{\cal A}(H^a,\bar H^a)}}{3(-{\rm i}(\tau-\bar{\tau}))(-\rmi(T-\bar T)+k(H^a,\bar H^a)f(U^A,\bar U^A)^{-\frac13})f(U^A,\bar U^A)},\\
c_2 &= \frac{e^{-4{\cal A}(H^a,\bar H^a)}}{3(-{\rm i}(\tau-\bar{\tau}))(-\rmi (T-\bar T)+k(H^a,\bar H^a)f(U^A,\bar{U}^A)^{-\frac13})^2f(U^A,\bar{U}^A)^\frac13},
\end{align}
$M^2 = \sqrt{2}$, and $\hat M_{ij}$ is given in \eqref{eq:MIJ}. The chiral multiplets $X$, $Y^i$, $P_L\Lambda_\alpha$ and $H^a$ describing the world volume fields are constrained as
\be
X^2=0,\qquad XY^i=0,\qquad XP_L\Lambda_\alpha=0,\qquad X\bar H^a=\text{chiral}
\ee
and contain, respectively, the Goldstino, the triplet of massive fermions, the U$(1)$ gauge vector and the three complex scalars as independent physical degrees of freedom. The scalar potential is
\begin{equation}
\mathcal{V} = e^K(|DW|^2-3 |W|^2) \supset V_{\Db}(H, \bar H) = 2e^{4\mathcal{A}(H,\bar H)-8u}
\end{equation}
and contains the anti-D3-brane uplift contribution leading to a de Sitter vacuum.

As a consistency check, one can verify that the anti-D3-brane supergravity action is invariant under ${\rm SL}(2,\mathbb{R})$ transformations, as it is expected in type IIB. The bulk fields transform as
\begin{equation}
\tau \to \frac{a\tau+b}{c\tau +d}, \qquad G_3 \to \frac{G_3}{c\tau+d},
\end{equation}
while the worldvolume fermions 
\begin{equation}
\label{fermtransf}
P_L\lambda \to e^{-\rmi \delta } P_L\lambda, \qquad P_L\chi^i \to e^{-\rmi \delta} P_L \chi^i, \qquad e^{-2\rmi \delta} = \left(\frac{c\bar \tau +d}{c\tau + d}\right)^\frac12.
\end{equation}
Then, one can check that the constrained multiplet $X$ and $Y^i$ have to transform as
\begin{equation}
X \to \frac{X}{c\tau +d}, \qquad Y^i \to \frac{Y^i}{c\tau+d},
\end{equation}
in order that the change in $K$ and $W$ amounts to a K\"ahler transformation
\begin{equation}
K \to K + \log|c\tau+d|^2, \qquad W \to \frac{W}{c\tau +d}.
\end{equation}
The part of the action containing the vector field is not invariant under ${\rm SL}(2,\mathbb{R})$ transformation, but is self-dual. This means that it has to be on-shell equivalent to an action of the same functional form, where the vector field is exchanged with its magnetic dual and the gauge kinetic function with its inverse. A detailed proof that this is indeed the case for the action of interest is given in \cite{Cribiori:2019hod}. It employs the fact that, due to the nilpotent constraint on $X$, $\hat f_{\Db}$ satisfies the property
\be
(\hat f_{\Db} (\bar f))^{-1} =\hat f_{\Db} (\bar f^{-1}).
\ee

\acknowledgments
I would like to thank F. Farakos, L. Martucci and T. Wrase for discussions. This work is supported by an FWF grant with the number P 30265.

\appendix
\section{Spinor conventions and dimensional reduction}
\label{appAferm}
A practical way to perform the dimensional reduction of the fermions is to use the explicit basis given in \cite{Bergshoeff:2015jxa}, whose construction is reviewed here.

Under the decomposition of the structure group ${\rm SO}(1,9)\rightarrow{\rm SO}(1,3)\times {\rm SO}(6)$, the ten-di\-men\-sion\-al gamma matrices split as $\tilde \Gamma^M = \{\tilde \Gamma^\mu,\tilde \Gamma^m\}$, where
\begin{equation}
\tilde \Gamma^\mu = \tilde \gamma^\mu \otimes \mathbb{I}_8,\qquad \tilde \Gamma^m = \gamma_* \otimes \tilde E^m.
\end{equation}
As in the main text, a tilde refers to tangent space quantities and the indices are $\mu=0,\dots,3$, $m=4,\dots,9$. The ${\rm SO}(1,9)$ chiral gamma matrix is
\begin{equation}
\Gamma_* = \gamma_*\otimes E_* = - \tilde \Gamma_{0123456789},
\end{equation}
where $\gamma_* = {\rm i}\tilde \gamma_{0123}$ and $E_* = {\rm i} \tilde E_{456789}$. These chiral gamma matrices are used to construct chiral projectors in the respective dimensions. The ten-dimensional charge conjugation matrix, $C_{(10)} = C_{(4)}\otimes C_{(6)}$, satisfies $C^T_{(10)} = - C_{(10)}$, while the lower dimensional ones are such that $C^T_{(4)} = - C_{(4)}$ and $C^T_{(6)} =  C_{(6)}$ respectively.

This decomposition is quite general. However, the idea in \cite{Bergshoeff:2015jxa} is to give a specific basis in which the identification of a subgroup ${\rm SU}(3)\subset {\rm SO}(6)$ as the holonomy group of the Calabi--Yau is straightforward. As a first step, one has to further decompose
\begin{equation}
\label{decSO1,9}
{\rm SO}(1,9)\rightarrow{\rm SO}(1,3)\times {\rm SO}(6) \rightarrow {\rm SO}(1,3)\times {\rm SU}(3).
\end{equation}
Recall now that in type IIB there is a doublet of Majorana--Weyl spinors $(\theta_1,\theta_2)^T$ of the same (positive) 10d chirality, $\Gamma_* \theta_{1,2} = \theta_{1,2}$, each of them transforming in the spinorial representation of ${\rm SO}(1,9)$, which is sixteen-dimensional. In the background under consideration, these two spinors are not independent, since in order to make the orientifold compatible with the kappa-symmetry on the anti-D3-brane, one poses a condition of the type
\begin{equation}
\theta_2 = \tilde \Gamma_{0123} \theta_1.
\end{equation} 
Therefore, it is sufficient to concentrate only one of them, which is $\theta_1\equiv \theta$ in the main text. Under the decomposition \eqref{decSO1,9}, one has then 
\begin{equation}
16 \rightarrow (2,4)\oplus(\bar 2, \bar 4) \rightarrow (2,3)\oplus(2,1)\oplus( \bar 2, \bar 3)\oplus( \bar 2, \bar 1),
\end{equation}
which means that the 10d spinor $\theta$ can be decomposed into four 4d Weyl spinors $P_L \chi^i$, $P_L\lambda$ and their conjugates
\begin{equation}
\theta \rightarrow \{P_L \chi^i,P_L\lambda,P_R \chi^i,P_R\lambda\}.
\end{equation}
These spinors are eigenvectors, with eigenvalues $\pm 1$, of the three commuting hermitean matrices
\begin{equation}
S^i = {\rm i}\tilde \Gamma^{i+3}\tilde \Gamma^{i+6},\qquad i=1,2,3,.
\end{equation}
In particular, the ${\rm SU}(3)$-singlet $P_L\lambda$ ($P_R\lambda)$ has only $+1$ $(-1)$ eigenvalues, while the ${\rm SU}(3)$-triplets $P_L\chi^i$, $P_R \chi^i$ have mixed $\pm 1$ eigenvalues. 
This SU(3) factor is then identified with the holonomy group of the Calabi--Yau. Following \cite{Bergshoeff:2015jxa}, one can then specify this discussion to the basis in which the matrices $S^i$ are diagonal. In such a basis, given in eq. (A.6) of \cite{Bergshoeff:2015jxa}, the four 4d spinors are then embedded into the 10d spinor $\theta$ as
\begin{equation}
\sqrt{2}\theta = \left(P_R \lambda, P_L\chi^3,P_L\chi^2,P_R\chi^1,P_L\chi^1,P_R\chi^2,P_R\chi^3,P_L\lambda\right)^T,
\end{equation}
where the factor $\sqrt 2$ is introduced to have a canonically normalized kinetic term in the action. The dimensional reduction of the fermionic action becomes now straightforward. For example, the term containing $F_1$ in \eqref{Sfer10d} gives
\begin{equation}
-\frac{e^\phi}{4}F_\mu \bar\theta \Gamma^\mu \tilde \Gamma_{0123}\theta = -\frac{i}{4}e^{\phi}F_\mu \bar \theta (P_L-P_R) \Gamma^\mu \theta = \frac i4 e^\phi \partial_\mu C_0 (\bar \lambda P_R \gamma^\mu \lambda + \delta_{i\bar\jmath}\bar \chi^{\bar\jmath}P_R\gamma^\mu \chi^i).
\end{equation}
Similarly, the term with $F_5$
\begin{equation}
\frac{1}{4!} \bar \theta \Gamma^{\mu npqr} F_{\mu npqr} \tilde \Gamma_{0123}\theta = \bar \theta \Gamma^{\mu i \bar \imath j \bar\jmath} F_{\mu i \bar \imath j \bar\jmath} \tilde \Gamma_{0123}\theta = -{\rm i}\frac{\partial_\mu {\rm Re}T}{{\rm Im}T}(3\bar \lambda P_R \gamma^\mu \lambda - \delta_{i\bar\jmath}\bar \chi^{\bar \jmath}P_L \gamma^\mu\chi^i),
\end{equation}
where $\Gamma^i$ and $\Gamma^{\bar\jmath}$ are the complexified gamma-matrices defined in \cite{Bergshoeff:2015jxa}. The other terms in \eqref{Sfer10d} can be calculated in an analogous manner.

\bibliography{refs}
\bibliographystyle{unsrt}

\end{document}